\documentclass{article}
\usepackage{graphicx,amssymb}
\usepackage[rightcaption]{sidecap}
\usepackage{natbib}
\usepackage[margin=0.5in]{geometry}
\usepackage{color}

\usepackage[english]{babel}

\usepackage{graphicx}
\usepackage[colorlinks=true, allcolors=blue]{hyperref}
\usepackage{natbib}
\usepackage{amsmath}
\usepackage{authblk}

\title{The SKA Galactic Centre Survey \\
A White Paper}
\author[1]{Rainer Sch{\"o}del}
\author[1]{Antx\'on Alberdi}
\author[2]{Izaskun Jim\'enez-Serra}
\author[3]{Farhad Yusef-Zadeh}
\author[1]{Angela Gardini}
\author[4]{Michael Kramer}
\author[1]{Miguel P\'erez-Torres}
\author[5]{Mark R. Morris}
\author[6]{Jan Forbrich}
\author[7]{Adriano Ingallinera}
\author[8]{Francisco Nogueras-Lara}
\author[9]{Jonathan D. Henshaw}
\author[9]{Steven N. Longmore}
\author[1]{Javier Mold\'on}
\author[10,11,12]{Ian Heywood}
\author[4]{Isabella Rammala}
\author[13]{Farideh Mazoochi}
\author[13]{Fatemeh Tabatabaei}
\author[1]{Lourdes Verdes Montenegro}
\author[1]{Susana S\'anchez Exp\'osito}

\affil[1]{Instituto de Astrof\'isica de Andaluc\'ia (CSIC), Granada, Spain}
\affil[2]{Centro de Astrobiolog\'ia (CAB), CSIC-INTA, Torrej\'on de Ardoz, Spain}
\affil[3]{Northwestern University, Evanston, IL, USA}
\affil[4]{Max-Planck-Institute for Radio Astronomy, Bonn, Germany}
\affil[5]{University of California Los Angeles, Los Angeles,  CA, USA}
\affil[6]{University of Hertfordshire, Centre for Astrophysics Research, College Lane, Hatfield AL10 9AB, UK}
\affil[7]{INAF - Osservatorio Astrofisico di Catania, Catania, Italy}
\affil[8]{European Southern Observatory (ESO), Garching bei M\"unchen, Germany}
\affil[9]{Liverpool John Moores University, Liverpool, UK}
\affil[10]{University of Oxford, Oxford, UK}
\affil[11]{Rhodes University, Makhanda, South Africa}
\affil[12]{South African Radio Astronomy Observatory, South Africa}
\affil[13]{Institute for Research in Fundamental Sciences (IPM), School of Astronomy, Tehran, Iran}

\date{\today}

\newcommand*\aap{A\&A}

\newcommand*\ao{Appl.~Opt.}

\newcommand*\apjl{ApJ}

\newcommand*\apjs{ApJS}

\newcommand{\Msol}{M$_{\odot}$}

\newcommand{\sgra}{Sgr\,A*}

\begin{document}

\maketitle

\section{Abstract}

 With its extreme density of stars and stellar remnants, dense young massive clusters, high specific star formation rate, intense radiation field, high magnetic field strength, and properties of the interstellar medium that resemble those in high redshift galaxies and starbursts, the Galactic Centre is the most extreme environment that we can observe in detail.  It is also the only nucleus of a galaxy that we can observe with a resolution of just a few milli parsecs. This makes it a crucial target to understand the physics of galactic nuclei and star formation, as well as the connection between them. It enables studies of a large number of otherwise rare objects, such as extremely massive stars and stellar remnants, at a well-defined distance, thus facilitating the interpretation of their properties. The Galactic Centre has been and is being studied intensively with the most advanced facilities. In this White Paper, we advocate for a large-area, multi-wavelength survey with the Square Kilometre Array of an area of about $1.25^{\circ}\times0.3^{\circ}$ (180\,pc$\times$40\,pc), centered on the massive black hole Sagittarius\,A* and for repeated deep observations of the nuclear star cluster over a decade, which will allow the community to address multiple science problems with a single data set.

\section{Science Goals}

 A survey of the Galactic Centre (GC) with the Square Kilometre Array (SKA) will lead to exquisite data on the nearest galaxy nucleus and probe the physics of the Milky Way's most extreme environment in unprecedented detail and sensitivity. It will enable us to address a large number of key science cases, such as:

\begin{enumerate}
        \item What is the structure and  formation history of the GC as revealed by its stellar remnants?
        \item Pulsars: How many pulsars are there in the GC and where are they located? Is there a population of millisecond pulsars (MSPs) at the GC that can serve to test General Relativity (GR)? Can the presence of MSPs explain the GC  $\gamma-$ray excess?
        \item Massive stars: What are the properties of the winds of massive main and post-main sequence stars? Is the current day Initial Mass Function (IMF) at the GC different than in the Galactic disk?
        \item What is the present-day star-formation rate at the GC?
         \item Can the "building blocks" of life form in GC molecular clouds? How complex can chemistry become in the GC?
         \item What are the properties and physics of the large-scale magnetic field in the GC?
         \item What is the origin of the magnetised radio filaments in the GC?
        \item What is the activity cycle of accreting Neutron Stars (NSs) and Black Holes (BHs) in binaries? Can we improve on our understanding of the radio-near infrared emission correlation in Low Mass X-ray Binaries (LMXBs)? What is the mass function of NSs and BHs in the GC? (This science question will require follow-up observations with extremely large optical telescopes.)
        \item Can we confirm or exclude the presence of Intermediate Mass Black Holes (IMBHs) at the GC?
        \item What are the properties of infrared extinction towards the GC? 
    \end{enumerate}

Before discussing these science questions more in detail, we will give a brief summary of our current knowledge about galactic nuclei and about the centre of the Milky Way.

\section{The nuclei of galaxies}

Supermassive BHs (SMBHs) with masses of $10^{6-8}$\,$M_{\odot}$ lie at the
photometric and kinematic centres of all major galaxies \citep[see
review by][]{Kormendy:2013nr}. In addition, Nuclear Star Clusters
(NSCs) are present at the centres of practically all galaxies with
stellar masses $>10^{8}$\,$M_{\odot}$. With masses of a few
$10^{5-7}$\,$M_{\odot}$ and half-mass radii of a few parsecs, NSCs are
the densest stellar systems in the local Universe. The growth and
possibly even the formation of massive BHs is linked to the
NSCs in which they live \citep[see review
by][]{Neumayer:2020jw}. Interactions between the stars in NSCs
and central BHs can give rise to phenomena such as tidal
disruption events and gravitational wave emission from inspiral
events \citep{Alexander:2017fk}. Finally, Nuclear Stellar Discs (NSDs), dynamically cold (i.e.\
flat) rotating stellar systems with radii of a few 100\,pc, have been
found to surround the NSCs in barred spiral galaxies
\citep[see][]{Gadotti:2020xq}. In these galaxies, Central Molecular Zones (CMZs) are a natural consequence of the inward transport of material towards the nucleus driven by the galactic bar. This influx of gas provides the fuel for star formation that forms nuclear stellar discs and rings.

Galactic nuclei are
the most extreme environments in the local Universe. The stellar
population in nuclei has been found to be the most metal rich of their
host galaxies, with mean metallicities reaching twice solar or higher
\citep{Bittner:2020qx,Schultheis:2021du,Nogueras-Lara:2022by}.
Galactic nuclei thus contain stellar structures and populations that
are distinct in their properties and formation histories from the
surrounding galactic disc and bars or bulges.  Moreover, the Interstellar Medium (ISM) in galaxy nuclei in the local Universe  is dense, warm and turbulent, more similar to the conditions 
found in starbursts and high-redshift star-forming regions than those found in $z=0$ galaxy discs. \citep[e.g.][]{Kruijssen:2013tq,Levy:2022qc}. Galactic nuclei are
therefore of fundamental interest for astrophysics.

\begin{figure}[!htb]
\center
\includegraphics[width=0.9\textwidth]{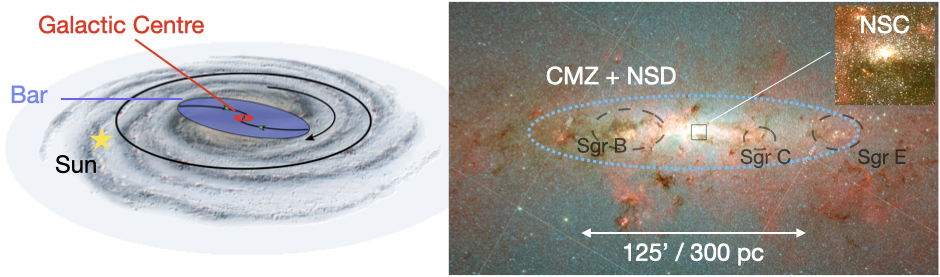}
\caption{\label{Fig:Galaxy} Left: Schematic view of our Galaxy. The Galactic
  Centre is indicated by the red circle, which outlines the location and size
  of the CMZ and NSD. Right: Spitzer $3.6 + 4.5 + 8.0\,\mu$m image of the GC (created from GLIMPSE data). The NSC surrounds Sgr\,A* and lies embedded in the NSD. The majority of the molecular gas in the CMZ is concentrated within the blue, dotted ellipse, which contains several prominent infrared  dark clouds. The NSD lies inside of the CMZ and partially overlaps with it. In addition to the NSC we have labelled three further regions of interest with dashed, grey lines: the Sgr\,B molecular cloud complex and star forming region \citep[including the  so-called dust ridge to its west, see][]{Henshaw:2022nm}; and the Sgr\,C and Sgr\,E HII regions.
}
\end{figure}

\section{The nucleus of the Milky Way \label{sec:MW_nucleus}}

The Milky Way is typical for large spiral galaxies in present-day low
density environments \citep{Bland-Hawthorn:2016qy}. It  is a barred
spiral with little evidence for any classical, spherical  bulge, which
is probably due to its quiescent merger history: All five of the up-to-now
identified significant merger events had stellar mass ratios smaller
than $1:31$ and occurred $>8$\,Gyr ago, with the exception of the
still ongoing Sagittarius merger \citep[see][]{Kruijssen:2020ev}.

The Milky Way plays a key role for astrophysics because it provides us
with the observational base for the vast majority of astrophysical
phenomena at the highest possible linear resolution and at the greatest
sensitivity. In particular, our galaxy is a unique laboratory for
understanding the nuclei of late-type galaxies.  The inward transport
of gas through the Milky Way's bar gives rise to its CMZ, which spans a region about  300\,pc across,  and
where up to 10\% of our Galaxy's molecular gas can be
found (Fig.\,\ref{Fig:Galaxy}). The conditions in this region \citep[and in similar regions observed
in the nuclei of nearby galaxies such as NGC 253, see][]{Sakamoto:2011lj,Martin:2021ot} are so extreme (in terms of density,
magnetic fields, turbulence, temperature) that the CMZ is our nearest
analogue of high-redshift star-forming regions
\citep{Kruijssen:2013tq}. 

The GC, is located at $8.25\pm0.01$\,kpc from Earth
\citep{Gravity-Collaboration:2020oy}. It is the only galactic nucleus in
which we can examine the properties and dynamics of stars on
milli-parsec scales \citep{Genzel:2010fk,Schodel:2014bn}. On scales of
tens to hundreds of parsecs, the GC is the most extreme environment in
the Milky Way and its most prolific star-forming region. It can serve
as a proxy to understand the conditions in CMZs in general \citep[e.g.][]{Kruijssen:2014ok,Henshaw:2022nm} and it is the best laboratory where to test whether key prebiotic precursors of the "building blocks" of life can form in interstellar space \citep{Jimenez-Serra:2020lo,Jimenez-Serra:2022sw}. Therefore, the GC is not only a template for galactic nuclei and for extreme astrophysical environments in general, but an essential piece to understand our cosmic origins.

Figure\,\ref{Fig:Galaxy} shows a sketch of the Milky Way and an overview of its central region. The GC can be considered to be outlined by the CMZ, which contains
$3-10\%$ of the Milky Way's molecular gas \citep[][]{Henshaw:2022nm} inside a surface area that is approximately 1000 times smaller than that of the Galactic Disc. The CMZ surrounds the Milky Way's NSD  and partially overlaps with it. The NSD is a flat, rotating stellar structure of about one
billion solar masses, with exponential radial and vertical scale
lengths of about  90\,pc ($\sim$38\,arcmin) and 30\,pc ($\sim$13\,arcmin) \citep{Launhardt:2002nx,Schonrich:2015uq,Schultheis:2019lw,Nogueras-Lara:2020pp,Sormani:2020ve}.

Averaged over the past 100\,Myr, the CMZ is our Galaxy's most
prolific star forming region \citep{Henshaw:2022nm}, with a mean specific star formation rate of about $1\times10^{-6}\,$M$_{\odot}$\,yr$^{-1}$\,pc$^{-2}$ (assuming a star formation rate of $0.1$\,M$_{\odot}$\,yr$^{-1}$ and a CMZ radius of 300 pc). For
comparison, assuming a Milky Way disc radius of 10\,kpc, the
present-day specific star formation rate of the entire Milky Way is
about $1\times10^{-8}$\,M$_{\odot}$\,yr$^{-1}$\,pc$^{-2}$, which is
100 times lower. The conditions in the CMZ
(large magnetic field, high temperature and turbulence of the ISM) may
favour the formation of massive stars \citep{Morris:1993ve}. The
detection of classical Cepheids and the analysis of stellar luminosity
functions suggest that on the order of $10^6\,M_{\odot}$ of stars
have formed in the GC in the past 20-30\,Myr
\citep{Matsunaga:2011uq,Nogueras-Lara:2020pp}. Even clusters more
massive than $10^4\,M_{\odot}$ appear to dissolve in $< 10$\,Myr due
to tidal evolution and tidal shocks in the GC environment
\citep[e.g.][]{Kim:1999db,Kim:2000cr,Portegies-Zwart:2002fk,Kruijssen:2014ok}. Currently, only two
massive young clusters are observed in the NSD, the Arches and the
Quintuplet cluster. Both have roughly $10^4\,M_{\odot}$, are 2-5\,Myr
old, and lie at about 25\,pc in projected distance from Sagittarius \,A* (Sgr\,A*).
Near-infrared \citep{Husmann:2012vn,Hosek:2019vn} and radio
\citep{Gallego-Calvente:2021yl,Gallego-Calvente:2022al} observations
of the Arches and Quintuplet clusters provide empirical evidence for a
top-heavy IMF (as compared to a Salpeter-like
function for massive stars in the Milky Way's disc).

The NSC dominates the inner $\sim$10\,pc of the
GC. With an effective radius of $4-5$\,pc (about two arc-minutes) and
a total mass of about $2.5\times10^{7}$\,M$_{\odot}$
\citep{Schodel:2014fk,Feldmeier-Krause:2017rt}, it is the densest and
most massive stellar system in our Galaxy. The exact relation between
the NSC and the NSD is not yet clear. But NSCs are almost omnipresent
in all types of galaxies \citep{Neumayer:2020jw}, while nuclear discs
or rings are preferentially found in barred galaxies
\citep{Gadotti:2019ig}.  Both the NSC and the NSD have supersolar mean
metallicities and are fairly old, but have different kinematic
properties and apparently different star formation histories
\citep{Schodel:2020qc,Schultheis:2021du}. The presence of massive,
young stars in the inner parsec of the NSC provides evidence for
in-situ star formation, probably in a previously existing accretion
disc, a few million years ago. Also, these young stars show a clearly
top-heavy IMF \citep{Bartko:2010fk,Lu:2013fk}.

The SMBH Sgr\,A* occupies the very
centre of the GC. Stars on short-period ($<20$\,yr) orbits around
Sgr\,A* have allowed us to measure its mass and distance precisely and
to perform first tests of GR in its surroundings
\citep{Gravity-Collaboration:2018qd,Do:2019ha,Gravity-Collaboration:2020oy}.

The GC region is permeated by a strong magnetic field that is
essential to understand the phenomena in this region, in particular the physics of
the ISM and star formation \citep[see][and references therein]{Yusef-Zadeh:1984dz,Morris:1996vn,Chuss:2003qe,Morris:2006uq,Ferriere:2009kx,Nishiyama:2010yn,Yusef-Zadeh:2022so,Butterfield:2024om,Lu:2024ey,Pare:2024uh,Tress:2024yy}. The GC magnetic field has two main
components, a poloidal one (often also called the vertical field), that appears to dominate the inner few 100
pc outside of the Galactic Plane, and a component that is found close to
the Galactic plane and is aligned parallel to the latter (frequently
called the horizontal field). The  vertical field is prominently
traced by non thermal filaments that can span tens of parsec in length
and are estimated to trace a field of $\sim$mG strength in the
intracloud ISM. The horizontal field is traced by much shorter thermal
filaments with field strengths a few tens to a
few 100$\mu$G, that are associated with denser gas and can show signs of deformation by the motion of the
ISM. 
 
A final point that needs to be mentioned when discussing the properties of the Milky Way's nucleus are the formidable observational obstacles. The dusty, molecular clouds along the line-of-sight to the GC cause strong attenuation of electromagnetic radiation in the optical to
ultraviolet regime ($\sim$$10^{-12}$ at visible wavelengths). Observations are therefore limited to the radio/mm, infrared, and
X/$\gamma$-ray regimes. Extreme source crowding requires
sub-arcsecond angular resolution to separate its stars and reliably identify counterparts of sources detected at different wavelengths via positional cross-matching. SKA1-Mid will provide us with angular resolutions $<0.1"$. It needs to be complemented by high angular resolution infrared imaging of the region, ideally by NIRCam/JWST \citep[see][]{Schoedel:2023rf}. Meanwhile, our best option is the GALACTICNUCLEUS survey that covers the inner $0.3$\,deg$^{2}$ of the GC with an angular resolution of $0.2"$ \citep{Nogueras-Lara:2018pr,Nogueras-Lara:2019yj}.

\section{The impact of the SKA on GC science}

\begin{figure}[!t]
\center
\includegraphics[width=\textwidth]{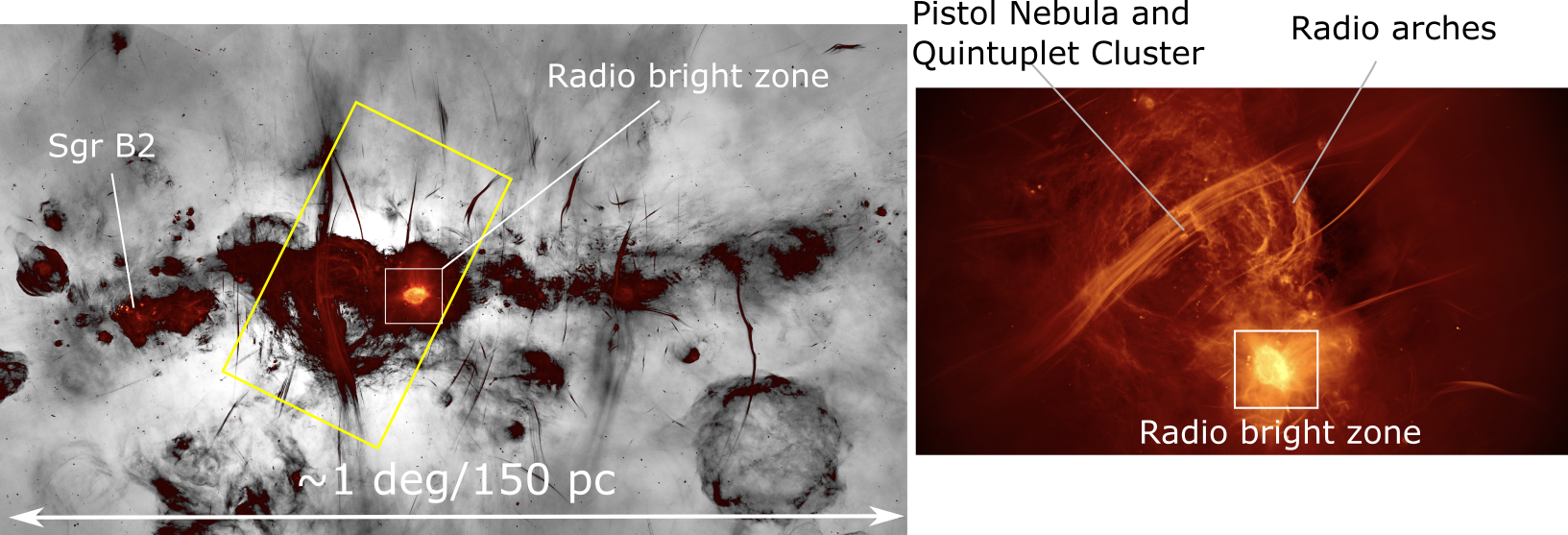}
\caption{\label{Fig:gc_radio} Left: MeerKAT $1.28$\,GHz image of the
  GC \citep{Heywood:2022rd}.  Galactic
  north is up, Galactic east is to the left.  Right: Zoom into the
  central area, that is indicated by the yellow rectangle in the left panel.
}
\end{figure}

The GC is a prime target for all major telescopes with rich synergies for multi-wavelength studies. This White Paper has a  strong emphasis on point-sources, i.e.~stars and stellar remnants. In this respect, SKA precursor instruments, such as
MeerKAT are  limited in their usefulness because of their insufficient angular resolution for the extremely crowded GC (e.g.~$\sim$$5"$ in the case of MeerKAT at $1.28$\,GHz). At the frequencies that we consider here, SKA1-Mid will provide us with a $\gtrsim10$ times higher angular resolution  at a significantly increased sensitivity (see Tab.\,\ref{tab:res-sens}) and will therefore enable a deep radio survey of a large field as well as repeated surveys of selected pointings for time-domain science and deep studies. The overall structure of the GC as seen in the radio regime is shown in Fig.\,\ref{Fig:gc_radio}. 

\begin{center}
\begin{table}[htb]
\centering
\begin{tabular}{lccc}
\hline
Band  & Central frequency  &  Beam  &  Continuum sensitivity\\
    &    [GHz]             &        &       [$\mu$Jy]  \\
\hline\hline
2 &  1.355 & $0.613"\times0.600"$  &  4.47 \\
5a & 6.55  & $0.127"\times0.124"$  &  0.74 \\
5b & 11.85  & $0.070"\times0.069"$  &  0.85 \\
\hline
\end{tabular}
\caption{Beam size and sensitivity of the SKA1-Mid in bands 2, 5a, and 5b. Estimated based on the \href{https://sensitivity-calculator.skao.int/mid}{SKAO Sensitivity Calculator}, assuming subarray configuration AA4 (15m antennas only), default bandwidths and central frequencies, 1\,h of on-source integration time, and Briggs weighting (Robust setting 0). \label{tab:res-sens}}
\end{table}
\end{center}

We continue to describe the key science cases and how SKA will impact them in more detail.

\subsection{\bf Stellar remnants and the structure and formation history of the GC} 
\label{sec:remnants}

The NSC and NSD can provide us with a proxy for the accretion history of the Milky Way's nucleus and thus about its evolution and probably also the growth of Sgr\,A*
  \citep[e.g.][]{Neumayer:2020jw,Nogueras-Lara:2020pp,Schultheis:2021du,Schodel:2020qc,Sormani:2022dq}. It
  appears that both the NSC and the NSD formed very early in our
  galaxy's history, at least $8-10$\,Gyr ago
  \citep{Nogueras-Lara:2020pp,Schodel:2020qc,Sanders:2022qa}. Their
  high masses and old ages imply the existence of large numbers of
  stellar remnants.  Stellar remnants, in particular NSs and BHs, can
  provide us with a window into past {\it massive} star formation and
  stellar dynamics, i.e.\ the formation of binaries that contain BHs or NSs. We can expect a large population of   LMXBs to form in the dense GC environment, because the relevant tidal capture mechanisms depend on the square of the stellar density \citep{Voss:2007kb}. 
  
  When LMXBs undergo transient accretion
  activity, they can be detected via their X-ray emission
  \citep[see][]{Muno:2005bh,Hailey:2018eq}. X-ray observations need to
  be carried out from space, which makes them expensive. Also, they provide an angular resolution of at best $0.5''$ (with the Chandra  X-ray observatory), which makes it challenging to pin down the  precise location of sources  and thus their potential stellar counterparts.  Fortunately, there is a tight correlation between X-ray and radio activity in LMXBs \citep[e.g.][]{Russell:2006da,Corbel:2013ie,Gallo:2018ap}, which makes radio observations an attractive option to infer the number and precise location of stellar remnants in the GC. With its great sensitivity, the SKA will be able to detect LMXBs even in rather faint states. Moreover, the SKA will provide the opportunity to 
  revisit these stellar remnants along their accretion activity cycle.

\begin{center}
\begin{table}[h]
\centering
\begin{tabular}{llll}
\hline
        &  WDs  &  NSs  & BHs\\
\hline\hline
NSD &  $7.2\times10^{7}$ & $3.5\times10^{6}$  &  $1.8\times10^{6}$ \\
NSC &  $2.6\times10^{6}$ & $1.3\times10^{5}$  &  $6.3\times10^{4}$ \\
\hline
\end{tabular}
\caption{Estimated number of stellar remnants in NSC
 and NSD. \label{tab:remnants}}
\end{table}
\end{center}

 We can obtain  estimates of the number of stellar remnants in the GC by assuming the star formation
histories of \citet{Nogueras-Lara:2020pp} for the NSD and of
\citet{Schodel:2020qc} for the NSC \footnote{We  approximate the respective
star formation histories with four single-age populations: (1) NSD:
90\% 8\,Gyr, 5\% 1\,Gyr, 4\% 250\,Myr, and 1\%\,40\,Myr, where the
percentages refer to the originally formed stellar mass at the given
age (not to the stellar mass still existing, which is lower). For its total mass we assume $7\times10^{8} \,M_{\odot}$
\citep{Sormani:2020ve}.  (2) NSC: 80\% 10\,Gyr, 15\% 3\,Gyr, 4\%
200\,Myr, and 1\%\,20\,Myr. For its total mass we assume
$2.5\times10^{7} \,M_{\odot}$ \citep{Schodel:2014fk}. Variations in
the exact ages will not affect in any significant way the number of NSs or stellar BHs, and
only to a small degree the number of white dwarfs (WDs). We used the
Stellar Population Interface for Stellar Evolution and Atmospheres
(SPISEA) Python package \citep{Hosek:2020mi} to simulate the stellar
populations and obtain the number of stellar remnants. We assumed
twice-solar metallicity \citep[for metallicities in the GC, see, for
example][]{Rich:2017rm,Nandakumar:2018zr,Nogueras-Lara:2020pp,
  Schodel:2020qc,Schultheis:2021du}. We used the {\it IFMR\_Raithel18}
class to assign initial-to-final mass ratios.  The fractions
of NSs and BHs originally formed in multiples is $\geq 0.99$ and those
of WDs formed in multiples is $\sim$$0.5$. For lack of any further information we assumed a power-law IMF constant with time (exponent $-2.3$ between 0.5 and 120 \Msol, $-1.3$ for smaller masses). \citet{Pfuhl:2011uq} find no evidence for a significantly altered IMF in the past. There is, however, significant evidence for a top-heavy IMF in the star formation event that created on the order of 200 massive stars within 0.5\,pc of \sgra about 4 million years ago \citep[see e.g.][]{Bartko:2010fk,Lu:2013fk}. Our estimate of the number of stellar remnants may therefore be conservative.}. Table\,\ref{tab:remnants} lists the estimated numbers of stellar remnants in the NSC and NSD. Different assumptions, such as the
number of star formation epochs, the uncertainties in the originally
formed stellar mass in each epoch, the metallicity, or initial-to-final mass ratios, may
result in changes by up to a few tens of percent, but will not impact the
overall magnitude of the numbers. Assuming a homogeneous distribution
of the remnants throughout the NSC and an approximate half-mass
radius of 4\,pc (1\,pc corresponds to about
$25''$ at the distance of the GC), a pencil-beam towards the NSC
with a radius of $1.0'$ may contain of the order of 15000 NSs, 7000 BHs, and
$3\times10^{5}$ White Dwarfs (WDs). These are conservative estimates, because the remnants are not expected to be
uniformly distributed, but mass segregated, with stellar BHs, in particular, significantly more concentrated towards Sgr\,A*
\citep[see][]{Morris:1993ve,Baumgardt:2018ad,Hailey:2018eq,Zhao:2022wb}. Even though we have not considered formation and evolution mechanisms of LMXBs here (which would go far beyond the purpose of this White Paper), these basic considerations illustrate how this large population of stellar remnants offers a great opportunity
to constrain the properties of these sources and the structure and
star formation history of the NSC and NSD.

The black hole low mass X-ray binary (BHLMXB) GX\,339-4 is one of the best-studied sources of its kind. Like the GC, it is located at a distance of about 8\,kpc. Its faintest and brightest observed radio luminosities lie between $0.05$ and $200$\,mJy \citep{Corbel:2013ie}, corresponding to a change of about 5 orders of magnitude in its X-ray luminosity.
The faintest reported observed radio flux of objects such as GX\,339-4 will therefore be accessible with the SKA1-Mid at high significance. Neutron Star Low Mass X-ray Binaries (NSLMXBs) behave in a similar way as BHLMXBs, but display, on average, about 20 times lower luminosity. This still implies that their faint hard states will be observable at the GC with the SKA1-Mid in 1\,h-long integrations (Fig.\,\ref{fig:sources}). In this context it is also interesting to point out the very faint and ultra-faint X-ray Binaries (XBs) detected at the GC and the possibility that  NSLMXBs may have recurrence times of decades between outbursts, much longer than often assumed \citep{Degenaar:2012yq,Degenaar:2015pt,Maccarone:2022kq}, which would imply that many of them may still not have been detected at the GC. Current X-ray transient surveys are limited to luminosities $\gtrsim1\times10^{34}$\,erg\,s$^{-1}$, corresponding to a few tens of $\mu$Jy at $\sim$10\,GHz for a BHLMXB located at the GC. With a $3\,\sigma$ sensitivity around $2.4\,\mu$Jy in 1\,h of observation at 11.85\,GHz, the discovery space of the SKA1-Mid will therefore be large: by pushing one order of magnitude deeper in the radio regime we can detect $\sim$40 times lower X-ray luminosities.

 The radio regime suffers practically no interstellar extinction, contrary to X-rays and the infrared. Observations of LMXBs will be key to address fundamental questions about the  structure and evolution of the Milky Way's nucleus.  Since the distribution of LMXBs will follow the underlying stellar density, their observation provides us with a way to determine the shape and symmetry of the NSD and NSC. This is, for example, of great importance for the NSC, because its southern part is hidden by large dark clouds \citep{Schodel:2014bn,Gallego-Cano:2020jq,Zhu:2018el}. Observations of BHLMXBs will also enable tests of mass segregation in the NSC \citep[e.g.][]{Morris:1993ve, Zhao:2022wb}. Once we know the distribution of stellar BHs
around Sgr\,A*, more accurate predictions can be obtained for extreme
mass-ratio inspiral events that may be observed in other galaxies with
the LISA gravitational wave observatory.
 
 Although it is currently assumed that the NSC and NSD are different structures, they may be related, with a smooth transition between them, as suggested by \citet{Nogueras-Lara:2023mi}. The ratio between stellar mass and the number of stellar remnants for the two structures may provide additional constraints on their relation \citep[but the effects of the dynamical formation of XBs may have to be taken into account][]{Generozov:2018cs}. 
 
  The number of LMXBs at the GC may be able to constrain whether massive stars formed preferentially at the GC, not just in the most recent star formation event, but also in the past (combined with star formation history and kinematic models; a challenging undertaking).

\smallskip
\noindent\underline{\it State-of-the-art.}
\citet{Zhao:2022wb} report on the detection of what they call hyper-compact radio
sources in $33.0$ and $44.6$\,GHz JVLA A configuration images of the
central $0.8$\,pc$\times0.8$\,pc ($20"\times20"$) of the Milky
Way. They define as ''hyper-compact'' those sources that have a major
axis size of $<0.1"$. After applying strict detection thresholds 
%of $S_{33.0}/\sigma_{\mathrm 33.0} = 15$
%($\sigma_{\mathrm rms, 33.0} = 8\,\mu$Jy\,beam$^{-1}$) in the Ka-band
%and and $S_{44.6}/\sigma_{\mathrm 44.6} = 10$
%($\sigma_{\mathrm rms, 44.6} = 17\,\mu$Jy\,beam$^{-1}$) in the Q-band
and requiring that the Ka- and Q-band positions must have a relative
positional offset $<3\,\sigma$, they identify 64 such sources. A large
sub-sample of 38 of those sources (58$\%$) have steep spectra,
$S_{\nu}\propto\nu^{\alpha}$ with $\alpha=-1.8\pm0.2$. These are
probably associated with LMXBs and are strongly
concentrated towards the central SMBH, as expected from theory and in
agreement with X-ray observations
\citep{Morris:1993ve,Muno:2005bh,Hailey:2018eq,Generozov:2018cs,Zhao:2022wb}.

\smallskip
\noindent\underline{\it Impact of the SKA.} 
SKA1-Mid will be able to reach angular resolutions of $0.13"$  and
$0.07"$ at $6.55$ and $11.85$\,GHz with an rms noise of
$0.7-0.8\,\mu$Jy\,beam$^{-1}$ in 1\,h integrations. At $11.85$\,GHz its angular resolution is comparable to the one  of the JVLA at $33$\,GHz, but 
with a much rounder beam, given the southern location of the GC. In addition, the high angular resolution of the SKA1-Mid will be
available at a significantly lower frequency, which favours the
detection of the type of steep-spectrum sources that are probably
associated with massive stellar remnants (synchrotron emission). 
%The SKA1-Mid would be about ten
%times more sensitive to flat-spectrum sources than the JVLA.  The
%search for hyper-compact radio sources at the GC with SKA1-Mid would
%also profit greatly from a $33-44$\,GHz receiver at the SKA1-Mid.
A more general discussion of the impact of SKA on the study of stellar remnants can be found in \citet{Corbel:2015fk}.

\subsection {Pulsars} 

Pulsars serve, among others, to perform fundamental tests of physics \citep{Kramer:2016uv}.  The study of radio pulsars at the GC is one of the central SKA science cases, because we expect a large number of them in this region and thus a high probability of finding double-pulsars or a pulsar in orbit around a stellar BH. A pulsar may even be present in a tight  orbit around Sgr\,A*. Both scenarios  would provide an exquisite probe of the physics of gravity \citep{Kramer:2004is,Eatough:2015fm}. Even though the probability of observing a normal pulsar (NP) on a short period ($\sim$1\,yr) orbit around Sgr\,A* that is beamed towards Earth may be rather low  \citep{Schodel:2020qc}, the intense star-formation activity in the GC in the past $\sim$10\,Myr suggests the presence of (up to a few) $100$ pulsars  throughout the central parsec \citep{Eatough:2015fm,Schodel:2020qc}. About 20\% of them may be beamed towards Earth.

Apart from NPs, which are young NSs with spin periods of one to a few $0.1-1$\,s there exist the so-called {\it MSPs}, old NSs spun up to reach
  fast rotational periods by accretion from a stellar
  companion. Given the old age of the stellar population in the
  GC as well as the large stellar density, such pairs of
  NSs and donor stars may  form dynamically in close
  encounters. MSPs are of great interest, because they can serve to obtain extremely  accurate dynamics, and therefore insights on the structure and  dynamics of the GC. If they are found near Sgr\,A* or orbiting a stellar BH, they can be exquisite probes of GR/theories of gravity. This is actually true for any pulsar in a sufficiently compact orbit, which is hence a strong motivation to search for pulsars in the GC \citep{Liu:2012zx,Psaltis:2016cu}.

 The detection of pulsars in the GC would also enable detailed studies of the ISM structure through spatial analysis of scattering screens \citep[][]{Bower:2014jv}.

Surprisingly, to this day only seven pulsars have been detected in the GC region, which we define in this paper to be the region inside the CMZ (see sec.\,\ref{sec:MW_nucleus}). Six of these pulsars all lie at projected distances $R>10'$ ($R>24$\,pc) from Sgr\,A* \citep{Deneva:2009fz,Schnitzeler:2016za,Wongphechauxsorn:2024dg}. The magnetar is at $R$$\sim$$2.5"$ ($R$$\sim$0.1\,pc). 
  
This so-called {\it missing pulsar problem} is a
 key puzzle of GC astrophysics \citep[e.g.][]{Torne:2021hm}. A possible explanation may be the challenge to observe pulsars in this environment: Typically, pulsars and
magnetars are identified by their pulses, but scatter
broadening makes this task difficult at the GC
 \citep[e.g.\ ][]{Eatough:2015fm,Torne:2021hm}. There may also exist
  deeper astrophysical reasons that could explain the dearth of
  pulsars. \citet{Dexter:2014sf} and \citet{Zhao:2022wb}, for example, discuss the intriguing hypothesis that most
NPs may be born as magnetars in the GC. Since the spin-down time
of magnetars is much shorter than the one of NPs, this
hypothesis could explain the missing pulsars problem at the
GC. 

The GC $\gamma$-ray excess was discovered by Fermi/LAT more than a decade ago \citep{Hooper:2011aq}. It is of great interest for fundamental physics, because it may be caused by the self-annihilation of dark matter particles at the GC \citep[e.g.][]{Salati:2014sy}. An alternative explanation for this excess of GeV radiation is the presence of a substantial population of MSPs at the GC \citep[e.g.][]{Abazajian:2011zw}. With the SKA we will be able to test this hypothesis with sufficiently deep observations and thus be able to settle a long-standing question.

%\citet{Zhao:2021at} provide some evidence that magnetars may be
%characterised by their radio spectrum, which would allow us to
%identify them with SKA multiband observations.

\smallskip
\noindent\underline{\it State-of-the-art}. To this date, seven pulsars have been confirmed in the GC, three of which appear to be magnetars \citep{Wongphechauxsorn:2024dg}. As concerns the search for candidates, \citep{Zhao:2020bd} analysed multi-epoch JVLA data at 5.5\,GHz. They report the detection of 110 compact
sources (size $<1"$). They estimate mean flux densities of $50\,\mu$Jy for NPs and $5\,\mu$Jy for MSPs at the distance of the GC. From their analysis (based on the observed luminosity function and non-variability on a timescale of 6 years) we can infer that on the order 20 of their sources are potential NPs.

\citet{Zhao:2022wb} analysed $33.0$ and $44.6$\,GHz JVLA A configuration images of the central $0.8$\,pc$\times0.8$\,pc ($20"\times20"$) of the Milky
Way. They reach an angular resolution of $\sim$$0.05"$ and an rms noise of $17$ and $8\,\mu$Jy\,beam$^{-1}$ at 33 and 44\,GHz. 
They discover a small number ($15\%$) of flat-spectrum
hyper-compact regions, which may be unresolved peaks of the ISM in
complex mini-spiral HII region. 
About a quarter of their detections,
$26\%$, have inverted spectra, which rise towards higher
frequencies. Even though a large fraction of the sources with inverted spectra are probably related to
stellar winds around massive stars, \citet{Zhao:2022wb} speculate, based on their
study of the spectrum of the magnetar SGR\,J1745--2900,
that some  of the inverted-spectrum sources may be magnetars.

\smallskip
\noindent \underline{\it Impact of the SKA.}  SKA will be instrumental in finding  pulsars at the GC.  SKA1-Mid is expected to reach a continuum rms noise of $5.3$, $0.7$, and $0.8\,\mu$Jy\,beam$^{-1}$  at $1.355$, $6.55$, and
$11.85$\,GHz  in one-hour-long integrations.  Assuming $S_{\nu}\propto\nu^{\alpha}$ and a spectral
index of $\alpha=-1.6$ \citep{Jankowski:2018oy}, NPs will have mean flux
densities of about $510\,\mu$Jy at $1.4$\,GHz, $40\,\mu$Jy at
$6.7$\,GHz, and $12\,\mu$Jy at $12.5$\,GHz at the distance of GC.  The
corresponding values for MSPs will be $51\,\mu$Jy at $1.4$\,GHz,
$4\,\mu$Jy at $6.7$\,GHz, and $1.2\,\mu$Jy at $12.5$\,GHz.  Hence, at
$6.7$\,GHz the SKA will supersede the previously mentioned $5.5$\,GHz JVLA observations by a factor of roughly five in sensitivity and angular resolution. If we
assume a spectral index of $\alpha=0.6$ for the inverted-spectrum radio sources (hypothetical magnetars), then SKA1-Mid will still be about five times more sensitive at $11.85$\,GHz than the JVLA at $33.0$\,GHz, while also providing a rounder beam at an equivalent angular resolution. MeerKAT is evolving rapidly and is already superseding the JVLA in sensitivity at low frequencies, but is limited its in high frequency capabilities.

In 10\,h of observation, an rms noise of
$\sim$$0.2\,\mu$Jy\,beam$^{-1}$ can be reached at $6.7$\,GHz, which implies the
possibility to detect MSPs at the $7\,\sigma$ level with SKA1-Mid.  Follow-up pulsar-timing, multi-wavelength  and polarisation observations will be essential for confirming the nature of the sources and potentially identifying magnetars via their spectral properties (see next section). 

With the help of VLBI observations, which will profit significantly from the inclusion of SKA1-Mid, we will be able to  measure the proper motions of NPs at the GC \citep[see][]{Bower:2015kq}, which can  provide us with upper limits on  how deep they are located inside the potential of
Sgr\,A*. Finally, in 100\,h, e.g.\ from stacking ten epochs of 10\,h
observations, MSPs can be detected in all three bands
with a significance of $>10\,\sigma$, but interstellar scattering may prevent identifying their pulses. Assuming scattering as observed for the GC magnetar \citep[see e.g.][]{Spitler:2014qf}, observing frequencies above 30 GHz may be needed.

\subsection{Massive stars, their winds and the IMF at the GC}

Averaged by volume, the GC is the Milky Way's most prolific star-forming region. It contains a large number of massive stars, for example in the central  parsec, in the Arches and Quintuplet clusters, as well as
  distributed throughout the NSD
  \citep[see][]{Clark:2021fj}. Since all these stars are located at a well-known, well-defined distance, the GC is  a convenient laboratory where we can study the still poorly known evolution of massive stars, in particular their mass loss.

  Obtaining an accurate census of young, massive stars at the GC is
  highly challenging, because of extreme interstellar extinction and stellar crowding \citep[see][]{Schodel:2014bn}. Massive post-main-sequence stars have have been detected via narrow-band infrared, radio,  and X-ray observations all over the GC \citep[e.g.\
][]{Mauerhan:2010kb,Dong:2011ff}. However, existing observations have
incomplete coverage of the GC, have not been performed
systematically, and would have missed O-main-sequence stars. An alternative way to find young, massive
  stars is to identify them via the emission from their strong,
  ionised winds and, possibly, circumstellar nebulae (depending on their evolutionary state). Thus, SKA can help to constrain recent star formation  at the GC and study mass loss via stellar winds and circumstellar nebulae \citep[e.g. Pistol star near the Quintuplet cluster,][]{Lang:2005zr}.

\underline{State-of-the-art.} \citet{Gallego-Calvente:2021yl} and \citet{Gallego-Calvente:2022al}
observed the Arches and Quintuplet massive young clusters at the
GC (located at about 25\,pc or 10\,arcmin in projection from Sgr A*)
with the A configuration of the JVLA at 6 and 10\,GHz, reaching
an rms noise of about $9\,\mu$Jy\,beam$^{-1}$ at 6\,GHz and
$5\,\mu$Jy\,beam$^{-1}$ at 10\,GHz, with an angular resolution of
$\sim$ 0.5".  They correlate the discovered compact sources with
young massive stars detected in the near-infrared and report the
detection of 18 radio sources associated with massive stars in the
Arches cluster and 29 in the Quintuplet cluster, thus increasing the
detected number of such stars in these clusters by factors two to three.
\citet{Gallego-Calvente:2022al} also use the data to derive mass loss rates, identify potential
binaries and study the mass loss rate variability by comparing
observations at two different epochs. Remarkably, they find that the
observed numbers of radio stars imply a non-standard, top-heavy or
bottom-light initial mass function in both clusters (as compared to
the typical Salpeter or Kroupa IMF found in other regions of the Milky
Way), in agreement with what has been found in near-infrared
observations \citep{Husmann:2012vn,Hosek:2019vn}.

\underline{\it Impact of the SKA.} Due to the inverted spectrum of the radio
emission from stellar winds, the SKA1-Mid, with its lower frequency
receivers, will provide us with only a  moderate improvement
in sensitivity for the study of radio stars, compared to non-thermal
sources. However, its significantly higher angular resolution (about a
factor of 5) will allow us to disentangle massive stars in high-density regions such as Arches, Quintuplet or central parsec, and to obtain significantly better astrometry for kinematic measurements. With 10\,h long observations at
$12.5$\,GHz, SKA1-Mid can reach an rms noise as low as
$0.4\,\mu$Jy\,beam$^{-1}$ and thus detect winds from all massive post-main-sequence stars and from main-sequence O-stars down to about
30\,$M_{\odot}$ in the GC, and provide
us with a more complete picture of recent (very) massive star formation at
the GC. It will also allow us to study the wind mass loss from
hundreds of stars that are located at a well determined distance, thus
providing a firm footing for studies of stellar-wind mass loss. Note that SKA1-Mid will simultaneously observe several hydrogen radio recombination lines together with the continuum emission, which will be used to constrain the internal physical structure and kinematics of massive stellar winds by means of 3D radiative transfer modeling (using the MOdel for REcombination LInes or MORELI code; Baez-Rubio et al. 2014; Martinez-Henares et al. 2023). This will allow us to investigate the launching mechanisms as well as the physical properties of the winds (mass, momentum, and energy) in a statistically relevant sample.

SKA1-Mid will also
  provide stronger constraints on the mass function of massive stars
  in the GC than what has been possible until now.
  It should be noted that the combination of short and long uv spacings with the SKA observations will facilitate the imaging of both extended and compact emission simultaneously, allowing the detection of point sources even within the diffuse emission (synchrotron or thermal free-free emission from star-forming regions), which would not be possible with other instruments. 

\begin{figure}[htb]
\center
\includegraphics[width=.8\textwidth]{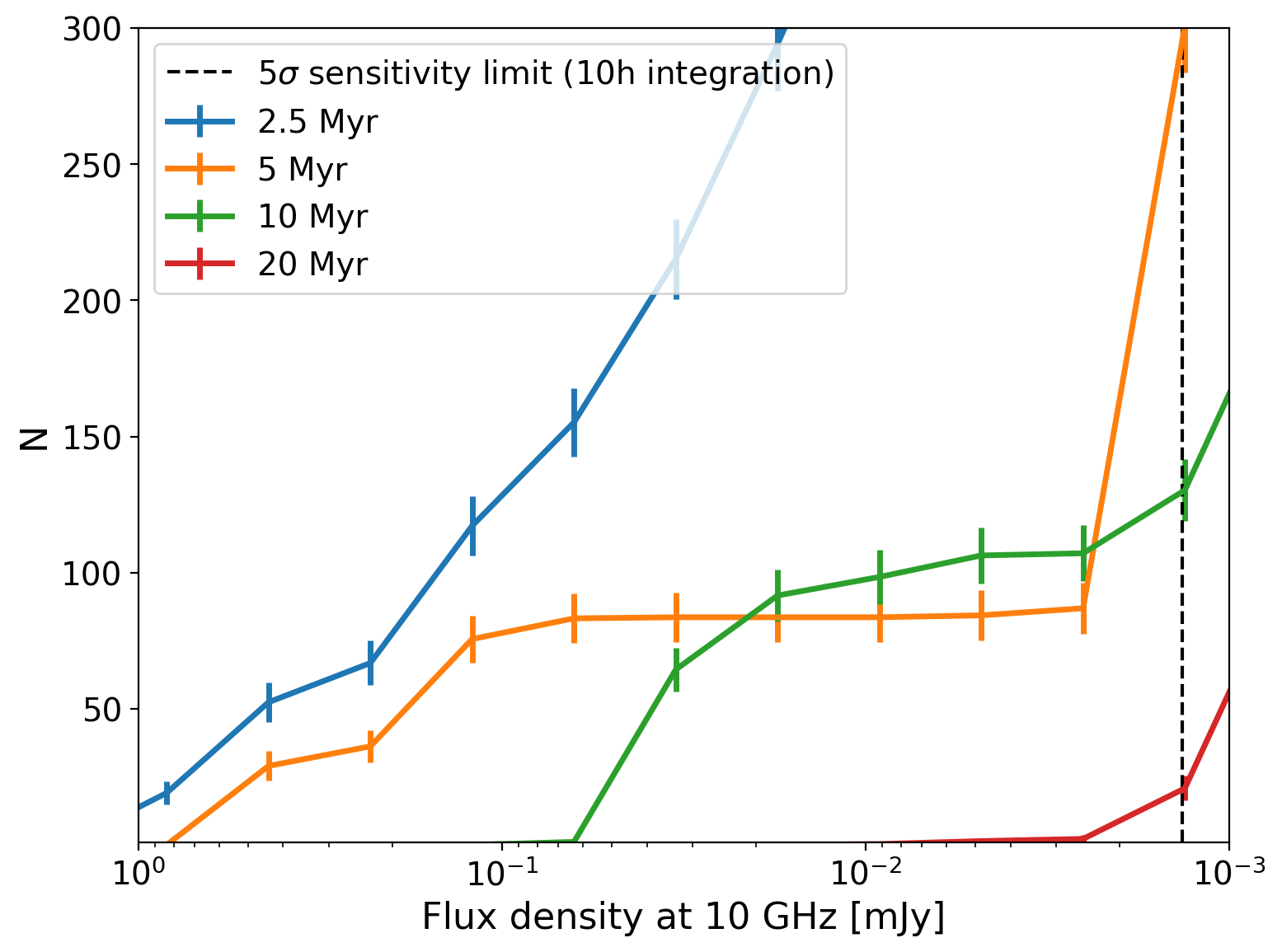}
\caption{\label{fig:RLF}  Cumulative radio luminosity functions for a star-forming event of $10^{5}$\,M$_{\odot}$ at the distance of the GC and with different assumed ages. Prediction from MIST isochrones, using Monte Carlo simulations to sample the mass function. The mass-loss rates of the stars that are provided by the MIST models were converted to radio luminosities \citep[e.g.][]{Leitherer:1997yf}. The RLFs have different characteristic shapes at each age, which can allow us to constrain both the age (via the shape of the RLF) and mass (total number) of  a star forming event.}
\end{figure}

Figure\,\ref{fig:RLF} shows the potential impact of studying the radio luminosity function of a region containing young stars. As an example we use the Sgr\,B1 region in the GC. It was recently reported that Sgr\,B1 contains about $10^{5}$\,M$_{\odot}$ of massive stars that formed roughly 10\,Myr ago \citep{Nogueras-Lara:2022jz}. Massive stars develop intense stellar winds once they turn off the main sequence. These winds give rise to thermal radio emission (with possible non-thermal contributions from colliding wind binaries). Radio emitting stars can be easily identified and distinguished from other radio sources by their necessarily precise astrometric coincidence with infrared-bright stars. With radio continuum observations we can measure the radio luminosity function of massive stars in a region such as Sgr\,B1. The radio luminosity function evolves significantly as a function of age, because it is a function of the rapid evolution of massive stars. Therefore radio continuum observations with SKA would be able to constrain the age of the star formation event that underlies the young population in HII regions such as  Sgr\,B1.

\subsection{Present day star formation} 

In spite of the  presence of copious amounts
  of molecular gas at extremely high densities, the Milky Way appears
  to be a significant outlier in the empirically derived relations between gas
  density and star formation activity \citep[see review
  by][]{Henshaw:2022nm}. One of the main difficulties in constraining
  the present-day star formation rate is that the extreme opacity of
  the molecular clouds at the GC may prevent us from
  detecting a significant fraction of the deeply-embedded 
  young stellar objects.  SKA1-Mid will cover a series of rotational lines of cyanopolyynes such as HC$_3$N, HC$_5$N, HC$_7$N and HC$_9$N \cite[see e.g.][]{Bianchi:2023fh}. These molecules are excellent probes of the deeply-embedded population of massive young stellar objects at the hot core stage\footnote{Hot cores are compact ($\leq$0.1 pc), dense (H$_2$ volume gas densities $\geq$10$^8$ cm$^{-3}$) and hot (T$\geq$150-300 K) condensations that represent the cradles of massive stars \citep{Garay:1999pr}.}. Indeed, they are largely enhanced in hot cores by the thermal evaporation of ices from grains as a result of the high temperatures of the dust \citep{Viti:2004to,Martin-Pintado:2005zt}. HC$_5$N is particularly interesting for SKA1-Mid because it presents four rotational lines in the v=0 state lying within bands 5a and 5b. The high-angular resolution of the proposed survey (beam of 0.07$"$- 0.13$"$) will filter out any low-excitation extended emission of HC$_5$N, pin-pointing the location of the high-excitation HC$_5$N emission arising from hot cores. Therefore, molecular line observations of HC$_5$N with SKA1-Mid (and, possibly, of even larger cyanopolyynes) will provide a complementary view to the one of the continuum images, giving information about the youngest population of massive stars in the making in the GC. 

 Last but not least, SKA1-Mid will also be able to probe maser emission from high-mass star formation (including methanol masers in the 6.7 and 12~GHz bands). These would not only be useful to pinpoint such regions of high-mass star formation, but additionally they could serve as a starting point for future SKA-VLBI astrometry and parallax measurements, along the lines of what the BeSSeL survey has provided for the northern sky (e.g., \citealp{Reid:2014jk}).

  \underline{\it State-of-the-art.} The line emission of HC$_3$N and HC$_5$N has recently been imaged across the CMZ within the ALMA Large Program ACES (ALMA CMZ Exploration Survey). This survey, however, has been acquired with an average angular resolution of 1.5$"$, insufficient to resolve the hot core population whose expected hot core sizes are $\sim$0.06$"$- 0.13$"$ (or $\sim$500 -1000 au at the distance of the GC).   
  
  \underline{\it Impact of the SKA.}  SKA1-Mid will cover all rotational lines from J=5 to J=2 in the ground vibrational level of HC$_5$N. HC$_5$N is measured to be factors of 2-3 less abundant than HC$_3$N in GC sources (see Zeng et al. 2018, 2018MNRAS.478.2962Z). For the hot cores of Sgr B2(N), the HC$_3$N column densities are of $\sim$10$^{17}$ cm$^{-2}$ within a beam of 1$"$ \citep[see][]{de-Vicente:2000sh}. These column densities are expected to be of $\sim$10$^{19}$ cm$^{-2}$ for hot core sizes of 0.1$"$. Therefore, we can assume a HC$_5$N column density of $\sim$3$\times$10$^{18}$ cm$^{-2}$ for the hot cores in the GC. Using this column density, a linewidth for the HC$_5$N emission of 8 km s$^{-1}$, and a gas temperature of 300 K \citep[typical of hot cores in the GC;][]{de-Vicente:2000sh}, we expect peak fluxes for the HC$_5$N rotational lines of $\sim$125-300 $\mu$Jy/beam. These lines will be detected with S/N$>$3-7 in integrated intensity, unveiling the population of GC hot cores. 
  
  %SKA can the outflows from YSOs throughout the GC and will thus
 % provide us with precise measurements of the current star formation
 % rate in this region.

\subsection{Searching for the ``building blocks'' of life}

The Giant Molecular Clouds in the GC are apparently the most chemically rich repositories of molecular material in our Galaxy \citep[e.g.][]{Requena-Torres:2008rw}. In the past few years, about two dozen new molecular species have been discovered in the dense clouds in the CMZ. Interestingly, some of these species are of prebiotic interest and include precursors of ribonucleotides, nucleobases, sugars, proto-proteins and proto-lipids \citep[e.g.][]{Rivilla:2021qc,Rivilla:2022mw,Rivilla:2023bd,Rodriguez-Almeida:2021ud,Rodriguez-Almeida:2021if,Zeng:2021kc,Sanz-Novo:2023hs,Jimenez-Serra:2020lo,Jimenez-Serra:2022sw}. These molecules are better detected in the CMZ because of the large column densities found in GC molecular clouds. Therefore, these clouds represent a unique opportunity to learn about how complex the chemistry of the ISM can become, and whether the building blocks of life could form already in interstellar space. 

 \underline{\it State-of-the-art.} All these molecules have been discovered in emission in the 7mm, 3mm and 2mm bands and they likely represent the tip-of-the-iceberg. However, due to the high sensitivity already reached by the single-dish, broadband spectral surveys already carried out with the Yebes 40m and IRAM 30m telescopes, we are starting to reach the line confusion limit in the observed spectra for these objects, which prevents the discovery of new prebiotic species of even higher complexity. The only way to circumvent this problem is to observe these molecules at even lower frequencies, which are much cleaner from simpler and smaller molecules. The problem is that the line transitions of large molecules such as prebiotic species become weaker since their partition functions become large (the population of any given molecule gets spread over a greater number of energy levels). In addition, at low frequencies their Einstein A$_{ul}$ coefficients become orders of magnitude smaller  (typically from 10$^{-6}$ s$^{-1}$ to 10$^{-9}$ s$^{-1}$). This issue is alleviated if we observe these species in absorption against strong background sources instead of in emission \citep{Jimenez-Serra:2022cr}. 
 
\underline{\it Impact of the SKA.} By performing spectroscopic surveys toward strong continuum centimeter sources, SKA1-Mid has the potential to discover new prebiotic interstellar species in absorption spectra. As an example, \citet{Jimenez-Serra:2022cr} have shown that a few rotational lines from the C3 sugar glyceraldehyde (i.e. a sugar with three carbon atoms) will be detected with a S/N$\geq$3 in integrated intensity with SKA1-Mid in bands 5a and 5b in 1 hour integration time. Larger sugars such as erythrulose (a C4 ketose sugar), however, will require hundreds of hours of integration time \citep{Jimenez-Serra:2022cr}. But the proposed survey will represent a key step for the discovery of large prebiotic molecules because it will reveal the location of the strongest continuum background sources in bands 5a and 5b toward which to carry out future dedicated high-sensitivity spectral surveys with SKA1-Mid.   

\subsection{Properties and physics of the large-scale magnetic field}

Among the biggest open questions concerning the GC magnetic field are:
(1) Is the milligauss vertical field homogeneous and pervasive
throughout the CMZ and does this define an overall GC
magnetosphere or are the strongest magnetic features only localised,
transient features? (2) How and where does the vertical field couple
to the horizontal field? Can we find specific points of interaction?
(3) What is the origin of the relativistic electrons that light up the
NTFs (see next subsection)? (4) What is the origin of the poloidal
field? 

 Studies of MeerKAT continuum
 data at 1.3\,GHz \citep[e.g.][]{Heywood:2022rd,Yusef-Zadeh:2023zt}
 have traced the thermal and non-thermal filaments with unprecedented
 sensitivity and  have provided an impressive new view of the GC
 magnetic field. Still, they  are limited to an angular resolution
 $\simeq 5^{"}$, which makes it hard to untangle different features and
 detect faint filaments in the radio bright GC.

 The SKA1-Mid in bands 2, 5a,
 and 5b will provide sensitive, high-resolution radio continuum maps,
 enabling to study the magnetic field with angular resolutions almost
 ten times better than the published MeerKAT observations. This will
 shed light on the role of the magnetic field in the GC on scales down
 to milli-parsecs. So far undiscovered faint filaments will improve
 our knowledge of the magnetic field and of  the origin of the
 relativistic electrons that power the NTFs. One particularly important
 question here is the existence or not of non-thermal horizontal filaments close
 to the Galactic plane. The data will provide us with a better
 understanding of the influence of the magnetic field on the dynamics
 of the ISM, which may even come to dominate some HII
 regions \citep[see][]{Bally:2024iu}. An exciting prospect is that the
 high angular resolution of SKA-MID may allow us to perform proper
 motion measurements of magnetic filaments over timespans of a few to
 ten years. Proper motions could help us clarify the relative locations,
 interactions and  whether the vertical field does or does not rotate with
 respect to the ISM and stars, which could indicate whether we are
 seeing global field configurations or whether the NTFs are
 localised, short-lived features.

\subsection{Origin of the non-thermal filaments}

One of the first hints showing that the nucleus of our Galaxy harboured energetic activity was the discovery of the archetype magnetised radio 
filaments in the GC 40 years ago \citep{Yusef-Zadeh:1984dz}. Since then, radio observations have shown a population of linearly polarised  synchrotron emission tracing nucleus-wide cosmic ray activity throughout the inner few hundred parsecs of the Galaxy \citep[e.g.][]{Heywood:2019fu}. 
Furthermore, H$_3^+$ absorption-line measurements of this region determined high cosmic-ray ionisation rates indicating that relativistic 
particles permeate the CMZ at levels a thousand times that in the solar neighbourhood 
\citep{Oka:2005ef,Indriolo:2012bv,Le-Petit:2016qf}. These particles provide a significant source of pressure in the GC when compared to 
thermal gas pressure in the ISM of the GC.  It is also becoming clear that there is another population of radio 
filaments detected in radio galaxies in poor clusters.   These extragalactic 
filaments show remarkably similar underlying physics to those of the GC, in spite of vastly different scale lengths and environments \citep{Yusef-Zadeh:2023zt}. The origin of this class of filaments is not understood either \citep{Ramatsoku:2020ei,Rudnick:2022on}.  However, 
the  similarities provide an opportunity to investigate the physical processes in the ISM of the GC and 
compare them to  the environment  of the  intracluster medium for the first time using detailed and sensitive SKA observations. 

One of the key questions is the origin of the filaments in the GC and in radio galaxies. There is no obvious source that powers these 
nonthermal filaments.  Although numerous models have been proposed to explain the origin of the filaments, there is no consensus how the 
filaments are produced.  One filament origin,  known as the cometary model,  invokes an energetic compact source, possibly a fast-moving pulsar or a mass-losing stellar wind sources, 
accelerating cosmic rays to high energies \cite{Yusef-Zadeh:2019pt}. Remarkably, motivated by the search for pulsation from the steep spectra compact source embedded within the snake filament \citep{Yusef-Zadeh:2024yi}, the recent discovery of the first GC MSPs \citep{Lower:2024jh}, likely associated with a filament $\sim$1\,arcmin from the snake filament (termed "Sunfish" filament by the author), motivates  further observations with the SKA  to determine if  pulsars 
power nonthermal radio filaments.

A search for compact radio sources associated with nonthermal radio filaments found an initial sample of 46 sources that lie near the ends of 
filaments using MeerKAT data \citep{Yusef-Zadeh:2023zt}. In the cometary scenario, the filaments result from the collection and draping of magnetic field lines by 
a moving stellar wind bubble or a pulsar with respect to the medium and forming a cometary tail.  The nature of these compact sources and their 
physical association with the filaments are not clear.  The SKA GC survey with its remarkable sensitivity, resolution and broad 
bandwidth will be able to test this model. In particular, the spectral index of the compact sources will be determined, placing constraints on 
steep spectrum pulsar candidates and stellar wind sources.  In addition, high spatial resolution will provide details of possible interaction of 
the compact radio sources with the filaments.  Lastly, the remarkable sensitivity will be able to uncover fainter filaments than those 
detected in the MeerKAT survey of the GC.  The brightness distribution of the filaments may have important implications on  
a process that involves the turbulent environment in which GC filaments are produced.

\subsection{Stellar remnants: activity cycles,  radio-infrared correlation, mass function}

Observations of the GC will enable studies of the properties of BH/NSLMXBs with a large sample at a well-known distance. They will allow us to improve significantly our understanding  of the infrared-radio-X-ray correlation of LMXBs \citep{Coriat:2009kx} with parallel infrared observations.
 Observations repeated over time scales of years will allow us to better constrain the recurrence cycles of BH/NSLMXBs.
With repeated observations over about a decade of operations, the SKA will  detect hundreds of LMXBs in the NSC and control fields. This will allow us to obtain better constraints on our so far poorly constrained knowledge about the recurrence time of outbursts in XBs. 
  
\citet{Russell:2006da} and \citet{Russell:2007jg} show that there
exists a tight correlation between X-ray and optical/near-infrared
emission for BHLMXBs (and probably also NSLXBs). SKA1-Mid can easily
detect and follow-up transients at the GC. Near-infrared cameras such as
ERIS/VLT can detect the near-infrared counterparts of BHLMXBs almost
down to quiescence. This offers the possibility to study the
radio-X-ray-infrared correlation of LMXBs (and thus accretion
physics) with coordinated observations with  high angular resolution near-infrraed imaging
instruments such as ERIS/VLT and, in particular, MOSAIC/ELT.

SKA1-Mid will allow us to find a highly complete sample of stellar remnants
in binary systems at the GC. Those systems may be followed up
spectroscopically and astrometrically by the ELT and the VLT
interferometer to determine the mass function of the remnants. This
can provide us with potentially hundreds of mass measurements in a
region with a well defined properties. Since the mean metallicity is
high (about twice solar at the GC), we can gain insights how
metallicity influences wind mass loss from massive stars and the final
mass of stellar BHs.

\subsection{Extinction towards the GC}

Interstellar extinction is extreme and highly variable as a function of the line-of-sight towards the GC \citep[see e.g.][]{Scoville:2003la,Nogueras-Lara:2020qn}. It is one of the greatest obstacles in our interpretation of infrared observations of stars in the GC. For example, accurate and precise measurements of extinction are needed to correct observed stellar magnitudes and thus to be able to classify stars approximately via photometric measurements \citep[e.g.][]{Schoedel:2023rf}. Also, the conversion of observed fluxes to absolute fluxes depends sensitively on an accurate knowledge of extinction, implying an insecurity of a factor of up to two in the estimated stellar masses of young, massive stars \citep[e.g.][]{Clark:2018yp}. Combining  measurements of radio hydrogen recombination lines with infrared HI emission lines is a highly reliable way of measuring absolute extinction as a function of wavelength towards the GC \citep[see][]{Scoville:2003la,Fritz:2011fk}. 

Emission from ionised ISM is pervasive in the GC. SKA observations of radio recombination lines, which are unaffected by extinction, can be combined with infrared recombination lines from  NIRCam/JWST \citep[see][]{Schoedel:2023rf}. Such data would allow us to determine the extinction curve with great accuracy across the GC and test whether and how it depends on the line of sight and on infrared wavelength \citep[see, e.g., discussion in][]{Nogueras-Lara:2021di}. In addition, knowledge of extinction towards different ISM features will allow us to constrain their relative distances along the line of sight and therefore the 3D-distribution of the ISM at the GC.

\subsection{IMBHs}

There have been speculations on the presence of IMBHs at the GC \citep[e.g.\ ][]{Tsuboi:2017qf}.  SKA1-Mid
would be sensitive enough to detect all potential IMBH candidates
throughout the GC, even when they are in a quiescent state, such as
Sgr\,A*. Their nature can then be inferred via variability studies, VLBI, and stellar kinematic studies.

\begin{figure}[!htb]
\center
\includegraphics[width=\textwidth]{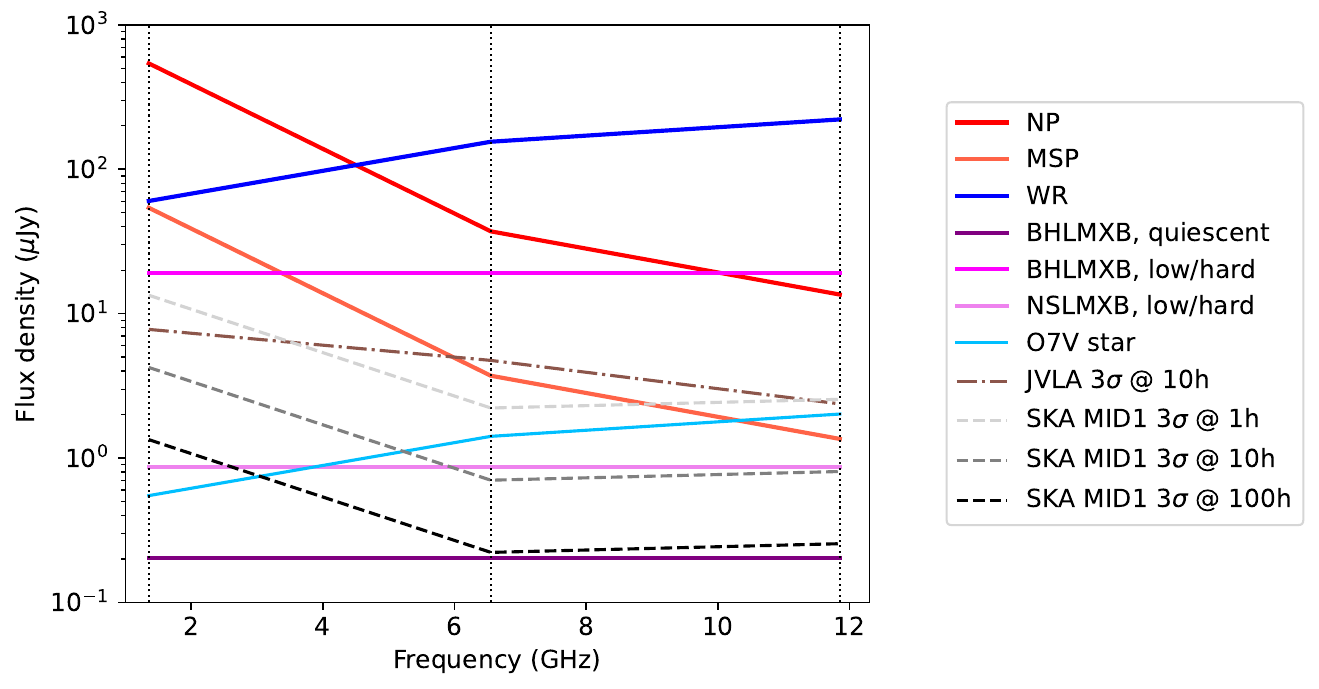}
\caption{\label{fig:sources} Flux densities of different types of
  radio point sources that we can expect to detect at the GC (solid lines). 
  Vertical dotted lines indicate the central frequencies of bands 2
  (1.355\,GHz), 5a (6.55\,GHz), and 5b (11.85\,GHz). The dashed gray to black
  lines indicate the $3\,\sigma$ rms of SKA1-Mid for
  integration times of 1\,h, 10\,h, and 100\,h. The dash-dotted brown line indicates the $3\,\sigma$ rms of the current JVLA. A GC distance of
  $8.25$\,kpc was assumed\citep{Gravity-Collaboration:2020oy}. NP: Normal pulsar,
  assuming a mean flux density of $50\,\mu$Jy at 5\,GHz and a spectral index of
  $-1.7$\citep{Zhao:2020bd}.  MSP: Millisecond pulsar,
  assuming a mean flux density of $5\,\mu$Jy  at 5\,GHz  and a spectral index of
  $-1.7$\citep{Zhao:2020bd}. WR: Wolf Rayet star, assuming a flux
  density of of $200\,\mu$Jy  at 10\,GHz  and a spectral index of
  $0.6$ \citep{Gallego-Calvente:2021yl}. O7V star: Thermal wind from an
  O7V star (approximately 30\,M$_{\odot}$) assuming a mass loss rate
  of $1\times10^{-7}$\,M$_{\odot}$yr$^{-1}$, based on a MIST version
  1.2 isochrone with solar metallicity and assuming a thermal spectral
  index of $0.6$. BHLMXB quiescent: Here we have used the observed flux density of the quiescent BHLMXB XTE J1118+480, observed by \citet{Gallo:2014ce}, scaling it to the distance of the GC and assuming a flat spectral
index. BHLMXB, low/hard: BH low mass X-ray binary in low/hard state,
assuming a flat spectral index and a radio/X-ray luminosity of
$10^{28}$\,erg\,s$^{-1}$/$10^{34}$\,erg\,s$^{-1}$. These fluxes correspond to the faintest detections given in 
\citet{Gallo:2018ap}, i.e.~we expect most sources to be brighter. NSLMXB, low/hard: NSLMXB in low/hard state,
assuming that NSLMXBs are about 22 times less radio loud than BHLMXBs \citep{Gallo:2018ap}. As for BHLMXBs, this is a lower limit.
%Many of these sources will be detectable with 1 hour of observing time, with the possibility of revisiting them in different observing runs in a very efficient way. It will also be possible to stack all  images, which will significantly improve the detection thresholds.
}
\end{figure}

\section{Properties of point-like radio sources at the GC}

Figure\,\ref{fig:sources} provides an overview of  point-like sources at the GC and their expected flux densities at different frequencies. Apart from their fluxes, the different types of point sources can be distinguished in various ways: Wolf Rayet and O-type main and post main sequence stars will have bright near-infrared counterparts and frequently thermal radio spectra. Pulsars will typically have steep spectra \citep[with the possible exception of magnetars, which may have flat to inverted spectra, see][]{Zhao:2022wb}, polarised emission \citep{Sobey:2022tn} and show little variability. LMXBs will show great variability with most of them manifesting as transient sources. In the low/hard state they typically show flat to inverted spectra from optically thick jet bases, i.e. $\alpha>0$ for $S_{\nu}\propto\lambda^{\alpha}$, where $S_{\nu}$ is the radio flux density and $\lambda$ the wavelength,  but they become optically thin ($\alpha<0$) in the high/soft state \citep{Fender:2004xv}. As concerns multi-wavelength cross-identification, Ultracompact HII and similar nebular sources  have sizes $\lesssim0.1$\,pc, corresponding to $\lesssim2.5"$ at the distance of the GC. Practically all of them will therefore be resolved and/or have an infrared counterpart. Near-infrared imaging with an angular resolution of $0.2"$ is available from the GALACTICNUCLEUS survey \citep{Nogueras-Lara:2019yj}.  Higher angular resolution imaging from JWST NIRCam has already been obtained on some selected fields (such as the central parsecs, Sgr\,C, or the Brick molecular cloud) and may become available for a significant fraction of the GC in the next years \citep{Schoedel:2023rf}. Also, specific fields may be observed  with adaptive optics assisted near-infrared instruments on large ground-based telescopes, such as ERIS/VLT.

\section{Survey requirements and possible setup}

We propose to study the GC  with a  survey that covers three aspects: (1) covering a wide field that encompasses the entire CMZ; (2) aiming at the highest sensitivity in the most extreme region in the GC,
the  NSC, that surrounds the
massive BH and probably contains a very  high concentration of
stellar remnants in the density cusp around the latter \citep[e.g.\ ][]{Baumgardt:2018ad}; (3) studying the time domain by repeated observations of fields in the NSC and NSD to constrain properties of stellar remnants, such as the recurrence time between LMXB outbursts, and massive stars, such as the variability of their winds, and to measure the proper motion of the point sources.

\begin{figure}[!b]
\includegraphics[width=\columnwidth,angle=0]{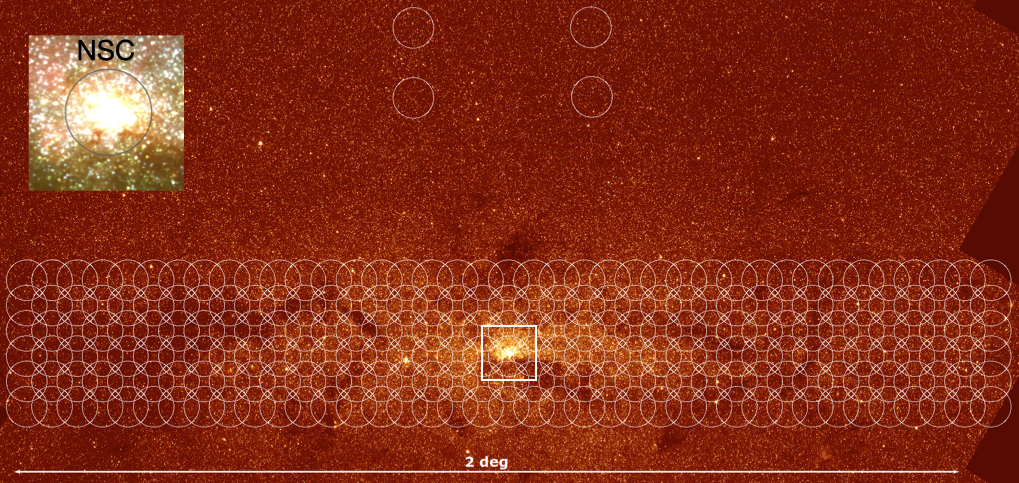}
\caption{\label{fig:pointings125} Possible layout of pointings at
  $12.5$\,GHz, overlaid over a Spitzer $3.6\,\mu$m image. Galactic
  north is up, Galactic east is to the left. The NSC is shown in the zoomed inlet, with the 12.5\,GHZ beam overplotted. The four disconnected circles to the Galactic north are the comparison fields.}
\end{figure}

Multi-wavelength observations will be necessary because spectral
indices provide us with insight into the nature of the radio emission
and therefore of the observed sources. The SKA GC survey should be
carried out at frequencies of $1.4$, $6.7$, and $12.5$\,GHz (bands
2, 5a, and 5b). The angular resolution at these frequencies will be $0.6", 0.13"$, and $0.07"$, respectively. This means that at the highest frequency and with a ten-year time baseline the proper motions of point sources detected at 10\,$\sigma$ can be measured with a precision  of 1\,mas\,yr$^{-1}$. The proper motions will serve to constrain to which stellar structure a source pertains \citep[e.g., see][]{Shahzamanian:2022vz} and how deep it is inside the gravitational potential at the GC.  In addition, polarisation measurements will be crucial to distinguish non-thermal sources as NSs or BHs from
other stars or thermal sources. Figure\,\ref{fig:sources} shows
 the flux densities expected from different objects of interest as a
 function of distance from Earth.

We assume an effective FWHM of the SKA1-Mid filed-of-view of $2.6'$ at
$12.5$\,GHz. Assuming 1\,h observations per pointing, we can cover a
$2.0^{\circ} \times 0.4^{\circ}$ field oriented parallel to the
Galactic Plane and centred on Sgr\,A* in 234\,h, as is shown in
Fig.\,\ref{fig:pointings125}. Additionally, several comparison fields should be observed in the inner Galactic Bulge so that we can assess the number and nature of sources that may overlap along the line of sight with the GC. We expect a relatively homogeneous source population at higher latitudes (mostly radio galaxies), therefore we estimate that four fields should be sufficient (4\,h per field and per band). This results in a time requirement of 250\,h.  Due to the larger field-of-view at the lower frequencies, the same field can be covered in about 70\,h at $6.7$ GHZ and 5\,h at $1.4$ GHz.  In total, we will require 325\,h for a GC survey that reaches an rms noise in the the continuum images of $2.0$, $1.3$, and $1.2\,\mu$Jy\,beam$^{-1}$ at $1.4$, $6.7$, and $12.5$ GHz, respectively. For the line images, the expected rms noise level will be 140, 90 and 85 $\mu$Jy\,beam$^{-1}$, respectively.  

This shallow survey should be complemented by repeated deep
observations of the NSC.
%, which is the astrophysically most interesting region of the GC. 
We propose to observe the NSC  with single 10\,h pointings in bands 5a and 5b. The sensitivity reached for each 10\,h pointing will be of $0.22$ and $0.26\,\mu$Jy\,beam$^{-1}$ at $6.7$, and $12.5$ GHz. 
These observations should be repeated at 30 different epochs over several years to allow us to observe variability and proper motions and to be able -- through stacking of the data -- to reach a sensitivity sufficient to detect quiescent XBs at $>5\sigma$ in band~5.
%The total time requirement will be 300\,h. 
%of which 100\,h at $12.5\,$, 30\,h at $6.7\,$ and 10\,h at $1.4\,$ GHz).
The  time requirement for the NSC study will thus be 600\,h.  In conclusion, we propose to invest roughly 925\,h for observing 
 the arguably most interesting environment of the
Milky Way with SKA1-Mid.

\section*{Acronyms}
\begin{tabular}{llll}
BH & Black Hole &  MSP & Millisecond Pulsar\\
CMZ & Central Molecular Zone & NS & Neutron Star \\
GC & Galactic Centre & NSC & Nuclear Star Cluster\\
GR & General Relativity & NSD & Nuclear Stellar Disc\\
IMBH & Intermediate Mass Black Hole & Sgr\,A* & Sagittarius\,A*\\
IMF & Initial Mass Function &  SKA & Square Kilometre Array\\
ISM & Interstellar Medium & SMBH & Supermassive Black hole\\
LMXB & Low Mass X-ray Binary & XB & X-ray binary\\
\end{tabular}

\section*{Acknowledgments} 
RS, AA, AG, MPT, JM, LVM, and SSE acknowledge the Spanish Prototype of an SRC 
(SPSRC) service and support funded by the Ministerio de Ciencia, 
Innovación y Universidades (MICIU), by the Junta de Andalucía, by the 
European Regional Development Funds (ERDF) and by the European Union 
NextGenerationEU/PRTR. The SPSRC acknowledges financial support from the
 Agencia Estatal de Investigación (AEI) through the "Center of 
Excellence Severo Ochoa" award to the Instituto de Astrofísica de 
Andalucía (IAA-CSIC) (SEV-2017-0709) and from the grant CEX2021-001131-S
 funded by MICIU/AEI/ 10.13039/501100011033.

\bibliography{/Users/rainer/Documents/BibDesk/BibGC}

\end{document}